\documentclass[preprint,1p]{elsarticle}
\usepackage{amsmath}
\usepackage{amssymb}
\usepackage{color}
\usepackage{graphicx}
\usepackage{multicol}
\usepackage{natbib}
\usepackage[utf8]{inputenc}

\journal{International Journal of Heat and Mass Transfer}

\begin{document}

\begin{frontmatter}
\title{The giant low-density lipoproteins (LDL) accumulation in the multi-layer artery wall model}
\author[sil,cebi]{Katarzyna Jesionek}
\author[sil,cebi]{ Marcin Kostur}
\address[sil]{Institute of Physics, University of Silesia, 40-007 Katowice, Poland}
\address[cebi]{Silesian Center for Education and Interdisciplinary Research, University of Silesia, 41-500 Chorzów, Poland}

\date{\today}

\begin{abstract}

The mathematical, four-layer model of the LDL transport across the
arterial wall including the sensitivity of the transport coefficients to the
wall shear stress (WSS) is studied.  In that model the
advection–diffusion equations in porous media are used to determine
the LDL concentration profiles in each layer of the arterial wall.  We
demonstrate the effect of the giant accumulation of the LDL in the intima
layer. This property turns out to be an interplay between the layered
structure of the arterial wall and WSS sensitivity of the endothelium.
Interestingly, we also show that the single-layer models with the same
WSS sensitivity mechanism and corresponding parameters, obscure the
accumulation phenomenon, which predicts the one order of magnitude larger
LDL concentration. 

The paper is supplemented by the repository with source code of the
model \cite{gitLDL}.

\end{abstract}


\end{frontmatter}

\section{Introduction}
Sudden cardiac death refers to estimated 4-5 million people per year
\cite{chugh2008epidemiology} and it is one of the major causes of mortality
in the adult population. Almost 80 \% of all sudden cardiac deaths is
connected with the myocardial ischemia caused by the atherosclerosis and more
specifically by coronary artery disease\cite{rosenberg2013height,
  chugh2004current}. For these reasons the pathogenesis of
atherosclerosis and evolution of atherosclerotic plaque are
extensively studied all over the world.

It is assumed that the development of atherogenesis is associated with
the abnormally high accumulation of the low-density lipoproteins in
the intima \cite{tarbell2003mass}. The LDL is responsible for the
transport of cholesterol from the liver to the all tissues of the
human body. Due to its putative role as an atherosclerosis precursor
the issue of the transport of the LDL molecules in the arterial walls
of the circulatory system and the factors influencing this transport
are widely studied.

One of such factors is the wall shear stress acting on the  wall of the
vessel, and directly on the endothelial cells. It is well accepted
that the occurrence of atherosclerosis is related to the exposure of
the endothelial cells to the low or oscillatory WSS
\cite{yiannis2007role}.

The impact of the wall shear stress on the LDL transport has recently
been a subject of various computational research. In order to properly
capture the processes occurring in the blood vessel wall, it is
necessary to create mathematical description of the complicated artery
wall structure. The class of so called 'multi-layer' models turned out
to reasonably well capture the properties of the arterial wall. An
example is the four-layer model proposed by Ai and Vafai in \cite
{ai_vafai_2006}.

In this work we study the properties of the arterial wall described as
a four-layer medium and, at the same time, sensitive to  WSS
value.

It is almost impossible to precisely obtain the transport properties
corresponding to the different layers of the wall during experimental
study. Therefore we will use two sets of estimated parameter values
reported in \cite{ai_vafai_2006, vafai2012}. The essential difference
between them is in the internal elastic lamina (IEL) properties. 

We will show that in the multi-layer models the LDL accumulation between
the layers with the low LDL permeability is possible. Depending on the used
set of parameters of the model, LDL concentration can be up to twenty times
larger then it was expected at the based on previously reported
results, e.g. \cite{olgac2008computational}.

This paper is structured as follows. Section \ref{suructure} describes
structure of the arterial wall with focus on the layers used in the
modeling.  Section \ref{LDLinWall} reviews known arterial wall models,
in particular the detailed description of the four-layer model,
developed by Ai and Vafai in \cite{ai_vafai_2006}.  In Section
\ref{effect_off_wss} we summarize the details of the wall shear stress
impact on the LDL transport, used in modeling. Section
\ref{computer_model} presents parameters and implementation of our
computer simulations.  In Section \ref{results} we discuss the results
of the simulations focusing on the giant LDL accumulation in the
intima.  Finally in Section \ref{summary} we will draw conclusions.

\section{Structure of the arterial wall} \label{suructure}

\begin{figure} 
\centering

\includegraphics [width=1.0\textwidth]{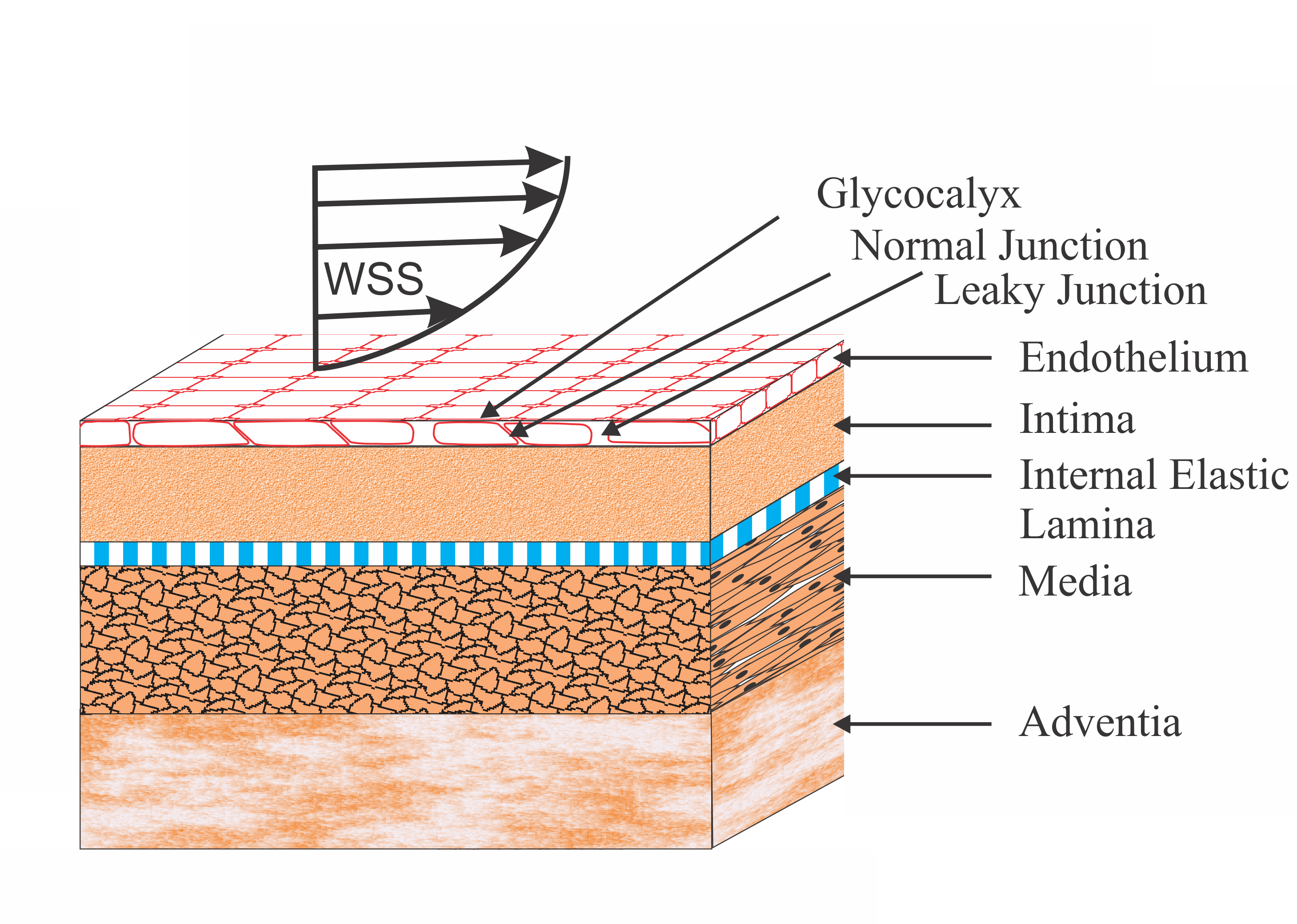}
\caption{Section of artery together with a schematic indication
of WSS profile near the wall.  }

\label{fig:StructBV}
\end{figure}

A typical large blood vessel has a layered structure, schematically
shown in the Figure \ref{fig:StructBV}.  It is composed of the six layers. The
glycocalyx is the first layer, which is in direct contact with
blood. Its thickness ranges from about $30$ to $100 nm$ (averaged
thickness is $60 nm$) \cite{michel1999}. It is a carbohydrate layer that
covers the  single layer of  endothelial cells and the
entrances of intercellular junctions \cite{vafai_yang_2006}.

Endothelium is the second $2\mu m$ thick layer.  It is a major barrier
in the LDL migration outside the lumen and at the same time has the
highest sensitivity to the conditions in the lumen. Endothelium changes
its transport properties depending on WSS value \cite{michel1999}. The
results contained in this paper are in large extend a consequence of
this sensitivity.

Directly behind the endothelium lies $10\mu m$ thick intima.  This is
a cushion layer made of connective tissue which may contain smooth
muscle cells and few fibroblasts. This layer is relatively easily
penetrable by both plasma and macromolecules \cite{vafai_yang_2006}.

Between the intima and the media is another membrane - internal
elastic lamina, constructed of impermeable connective tissue with
fenestral pores. Its thickness is assumed to be $2 \mu m$.  This
membrane is a significant barrier to the LDL lipoproteins. In this work we
will demonstrate that under certain conditions, this might be a cause
for the LDL accumulation \cite{vafai_yang_2006}.

Media, the thickest layer ($200 \mu m$), is made up of the alternating
layers of the smooth muscle cells and elastic connective tissue.  In media the
absorption of the LDL particles takes place\cite{vafai_yang_2006}.

Media is surrounded by the adventitia, which consists of collagen
fibres attaching the arterial wall to surrounding flabby connective
tissue.

\section{Modeling macromolecules transport in the arterial wall} \label{LDLinWall}

Realistic modeling of WSS effects on the LDL transport is still a big
challenge. Models including more details of the wall structure
obviously tent to be more difficult to implement and more demanding
computationally. At the same time they have more free parameters which
need to be precisely measured.  On the other hand oversimplified models
might overlook some important properties. Below we briefly review
commonly used approaches.
 
Prosi et al. \cite{prosi} divided vessel wall modeling into three
groups according to its degree of complexity. The simplest model
used i.a. in \cite{rappitsch1996pulsatile, wada2000computational} is
the wall-free model. It treats the arterial wall as a simplified
boundary condition. The advantage of this approach is its simplicity,
but that model does not allow to obtain information relating to the
transport in the arterial wall, but only in the lumen
\cite{olgac2008computational}.

More complicated, and thus more realistic is the lumen-wall model, in
which the transport of macromolecules is considered both in the lumen
and in the wall. This model was proposed by Moore and Ethier
\cite{moore1997oxygen}. Olgac et al. used this model in
\cite{olgac2008computational}, where the dependence of the LDL transport
properties on WSS was proposed. This model treats the arterial wall
as a homogeneous porous media. Hence, it is possible to observe
the macromolecules transport in the lumen and in the artery
wall, what is the main advantage of this approach. Despite of that,
this model is relatively simple in comparison to the last branch of
transport models, which is a multi-layer model.

Multi-layer model takes into account the internal structure of the
blood vessel wall, thus better reflects the properties of the actual
arterial wall. On the other hand this advantage causes the complexity
increase. One of the multi-layer models is the four-layer model proposed
by Ai and Vafai in \cite{ai_vafai_2006}. In this model, the arterial
wall is represented by the four layers: endothelium, intima, IEL and
media. Detailed description of this model is presented in the next
section.

\subsection{Four-layer model} \label{four_layer_model}

Ai and Vafai in \cite{ai_vafai_2006} developed an artery model taking
into account the layered structure of the blood vessel. Endothelium,
intima, IEL and media are the layers that are covered in this
model. They are shown in the Figure \ref{fig:StructBV}. Due to its
very small thickness, the impact of the glycocalyx is
neglected. Adventitia is taken into account as a boundary condition at
the opposite side of the system.

Endothelium and IEL are semi-permeable membranes and filtration theory
is applied to model them, while the intima and media can be modeled as a
porous media. The phenomenon of the osmosis for semi-permeable membranes
can be ignored, because the gradients of the LDL are too small to
influence the hydraulic processes \cite{vafai_yang_2006}. In this
case, the equations for transport through the membranes are effectively the
same as those for transport through the porous media, but with the
different coefficients for each layer. For this reason, the layers are all
treated as a macroscopically homogeneous porous media and mathematically
modeled using the volume averaged porous media equations, with the
Staverman filtration and osmotic reflection coefficients. These
coefficients allow to account for selective permeability of each
porous layer to the LDL macromolecules.

Finally, a description of the LDL transport in each layer of the
arterial wall is reduced to the following equations for the hydraulic
and filtration processes:

\begin{eqnarray}
\label{eq:pde1}
      \frac{\rho}{\delta}\frac{\partial\vec u}{\partial t} + \frac{\mu_{eff}}{K}\vec u & = & -\nabla p +\mu_{eff} \nabla^2\vec u\\
\label{eq:pde2}
      \frac{\partial c}{\partial t} +(1-\sigma)\vec u\cdot\nabla c & = & D_{eff} \nabla^2 c - k c
\end{eqnarray}
where $\rho$ is the fluid density, $\delta$ is the porosity, $\mu_{eff}$ is
the medium effective dynamic viscosity for the LDL, $K$ is the permeability, $p$ is
the pressure, $\sigma$ is the reflection coefficient, $\vec u$ is the  hydraulic
velocity of the solvent penetration and $D_{eff}$ is the diffusivity. The
model assumes first order decay of the LDL in media with the reaction rate
coefficient $k$.

Equation \ref{eq:pde1} in practice, can be replaced by the adiabatic
approximation in which the filtration velocity is determined immediately
and given by:

\begin{eqnarray}
	\label{eq:filtr}
	\vec u & = & - \frac{K}{\mu_{eff}}\nabla p 
\end{eqnarray}

The Equation \ref{eq:filtr} has a similar form to the (continuous) Ohm's
law for electrical circuits: $\frac{K}{\mu_{eff}}$ is the equivalent
of conductance, $\Delta p$ voltage, and $\vec u$ current. For this
reason, to the description of the connected layers, the analogy to
parallel and serial connections can be applied. Subsequent layers are
treated as resistors connected in series and the individual normal and
leaky junctions in the endothelial layer are treated as parallel
resistors.

\subsection{One-dimensional approximation} \label{1D}

System under consideration has very diverse scales. This allows to
express the hypothesis that major simplifications will not
substantially change the results of calculations. First, the wall is
thin compared to the length or circumference of the vessel. For
example, in the case of the coronary arteries, circuit is in range of
$ 10^4 \mu m$ and substantial thickness of the arterial wall does not
exceed $10^2 \mu m$. In addition, permeation processes are dominated by the
pressure forced filtration and predominantly occurs in the radial
direction. Thus the process of the LDL transport in the artery takes
place essentially in the this direction. As a result, our modeling
can be reduced to the one spatial dimension.

In the literature, a comparison of the two-and three-dimensional
models with the one dimensional ones confirms this assumption
\cite{vafai2012,vafai_yang_2006, vafai2008}. Additionally, we have
compared our one-dimensional results with three-dimensional simulation
done by Olgac et al. \cite{olgac2008computational}. The results are
discussed in the Section \ref{results}.

\section{Effect of WSS on LDL transport in the arterial wall} \label{effect_off_wss}

It has been observed, that atherosclerotic lesions are formed in the
specific areas of the circulatory system: the outer wall of
bifurcations and the inner wall of curvatures. A common feature of
these regions is the presence of disturbed flow \cite{yiannis2007role,
  vanderlaan2004site}. That tendency was first noticed by Caro et al.
\cite{caro1969arterial}.  Whereas later it has been confirmed in
several different ways: by using the computational fluid dynamic
simulations \cite{asakura1990flow, ku1985pulsatile, moore1994fluid},
in-vivo studies in animals \cite{buchanan1999relation, cheng2005shear,
  gambillara2006plaque} and in vivo investigations in humans thanks
combination of the advanced medical imaging techniques and computer
simulations \cite{stone2003effect, wentzel2001relationship,
  wentzel2005does}.

For simple blood vessels the WSS amplitude ranges from 1.5-7 Pa and
has a positive time average \cite{yiannis2007role,
  malek1999hemodynamic, stone2003prediction,
  gimbrone2000endothelial}. Such physiological level of the shear
stress appears to play a protective role for the functional integrity
of the endothelial cells \cite{dimmeler1996shear}. However, in the
areas where the disturbed blood flow occurs, the low WSS is
observed. This WSS is defined by low amplitude, on average 0.02-1.2 Pa
\cite{yiannis2007role, stone2003effect, malek1999hemodynamic,
  gimbrone2000endothelial, frauenfelder2007flow}. In the case of low
WSS increased endothelial cell turnover rate is observed
\citep{stone2003effect}.

\subsection{WSS dependent mechanism of the LDL transport by the endothelium}  \label{leaky}

The mechanism of the macromolecules transport in the arterial wall is
in large part defined by the endothelium. This layer causes increased
hydraulic and mass transfer resistance.  It is related to the small
pore size. Therefore, factors which cause an increase in the effective
pore size have a significant impact on the flow in the endothelium,
and thus in the entire wall.

The endothelium cells are connected to each other by the intracellular
junctions. These junctions can be normal or leaky. Therefore, in this
layer three pathways can be distinguished: vesicular pathway, normal
junctions, and leaky junctions.  The vesicular pathway is responsible
only for ($\sim 9\%$) of the LDL transport\cite{olgac2008computational}
and is not taken into the account in this paper.
 
Intercellular junctions would not normally allow any significant
passage of the LDL even through breaks in the normal junction, because
the wide part of the cleft is expected to be of the order of LDL
molecule size. The average radius of the normal junction is $5.5 nm$,
which is smaller than the radius of the LDL molecule ($r_{m} =11
nm$). The large, leaky pores are associated with cells that are in the
process of cell turnover: either cell division (mitosis) or cell death
(apoptosis). The cause of the leaky junctions formation in these cells
is the weakening of the junctions during the process of division or
sloughed off this cells by the healthy neighbors
\cite{tarbell2003mass}.

Therefor the process consists of the flux of the solvent through
normal and leaky junctions \cite{olgac2008computational} with velocity
which has to be calculated in the model. Then, this velocity drives
the passage of LDL molecules through the leaky junctions. The key
role in this scheme, plays a fraction of leaky junctions $\phi$ in the
wall and its dependence on WSS.

It turns out that processes of apoptosis and mitosis are affected
by wall shear stress. Several studies have shown that a reduction in
the rate of blood flow or shear stress induces an increase in the rate
of endothelial cell apoptosis, whereas an increase in shear stress
exerts the opposite effect \cite{dimmeler1996shear, tarbell2003mass}.

\begin{figure} 
	\centering \includegraphics [width=1.0\textwidth]{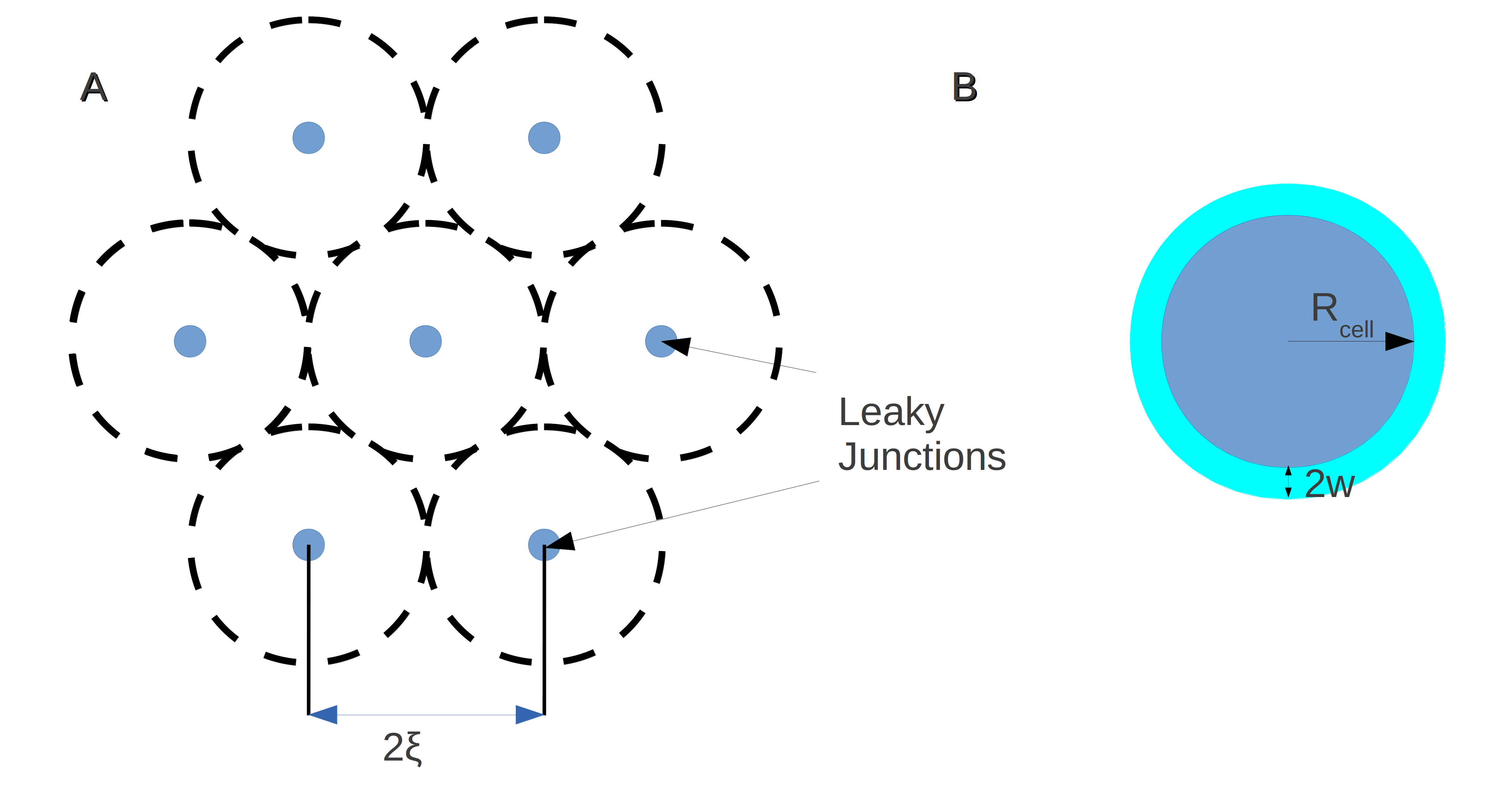}
	\caption{illustration of randomly distributed leaky cells. (A)
          A leaky cell represented by blue circle is present at the
          center of each periodic circular unit of radius $\xi$
          represented by dashed lines. (B) Leaky cell of radius
          $R_{cell}$ with leaky junction of half-width $w=10 nm$.}
	\label{fig:leaky}
\end{figure}

The fraction $\phi$ is defined as the ratio of the area of leaky
cells to the area of all cells \cite{dabagh2009transport}. This is shown 
schematically in the Figure \ref{fig:leaky}. According to this:
\begin{eqnarray}
	\label{eq:PHI}
    \phi=\frac{R_{cell}^{2}}{\xi^{2}}
\end{eqnarray}
where $R_{cell}$ is the radius of the endothelial cell taken as 
$15 \mu m$ and $\xi$ is radius of periodic circular unit
\cite{huang1994fiber}.

Olgac et al. in \cite{olgac2008computational} developed a model which
allows to calculate the fraction of leaky junctions. For self
integrity of the paper we will shortly review details of this approach.

The procedure was done in four steps, which we will summarize below.

In the first step the relation of the shape index (SI): 
\begin{equation}
	\label{eq:SIdef}
    SI=\frac{4\pi \times area}{(perimeter)^{2}}
\end{equation}  
and WSS was established. Based on experimental results from \cite{SI1, SI2, SI3} it was found that:
\begin{equation}
	\label{eq:SI}
    SI=0.380\times e^{0.79 \cdot \mathrm{WSS}}+0.225\times e^{0.043 \cdot \mathrm{WSS}}
\end{equation} 

In the second step the shape index was associated with the number of
mitotic cells per area of $0.64 mm^2$. The function fitted by Olgac et
al.\cite{olgac2008computational} to the data from\cite{MC} takes the
following form:

\begin{equation}
	\label{eq:MC}
    N_{MC}=0.003797 \times e^{14.75 \cdot SI }
\end{equation}

The third step was to get the ratio of leaky cells associated with
mitotic cells to all mitotic cells.  On that basis of the work
\cite{LC} Olgac et al. derived the correlation of the number of
mitotic cells with the number of all leaky junctions per area of $0.64
mm^2$. They assumed that the number of leaky cells correlated with
nonmitotic cells is independent on WSS. This relation is the
following:

\begin{equation}
	\label{eq:LC}
    N_{LC}=0.307+0.805\times N_{MC}
\end{equation}

In the last step, knowing the number of leaky cells in a given area
(in this case $0.64 mm^{2}$) fraction of leaky junctions was
determined to be:
\begin{equation}
	\label{eq:PHI_WSS}
    \phi=\frac{N_{LC} \times \pi R_{cell}^{2}}{6.4 \times 10^{-7}}
\end{equation}

The fraction of leaky junctions determine transport properties of the
endothelium. Since in the endothelium is modeled as a medium with
given diffusion $D$, reflection coefficient $\sigma$ and permeability
$K$ we need the quantitative dependence of those parameters on $\phi$.

The LDL molecule diffuses in cylindrical pores (effectively
treated as slits of width $2w$), therefore its diffusion coefficient can be
approximated by \cite{vafai2012}:
\begin{equation}
    D_{end}=D_{lj}=D_{lumen}(1-\alpha_{lj})F(\alpha_{lj})\frac{4w}{R_{cell}}\phi
    \label{eq:D_WSS}
\end{equation}
where $w$ is half width of leaky junctions equal $14 nm$ (see Figure
\ref{fig:leaky}). $\alpha_{lj}$ is the ratio of $r_{m}$ to $w$ and
$r_{m}$ radius of LDL molecule equal $11 nm$. $(1-\alpha_{lj})$ is the
partition coefficient, which is the ratio of the pore area available
for solute transport to the total pore area \cite{michel1999microvascular}:
\begin{equation}
    \alpha_{lj}=\frac{r_{m}}{w}
\end{equation}
F is a hindrance factor for diffusion in a pore, given by Curry
\cite{michel1999microvascular}:
\begin{equation}
F(\alpha_{lj})=(1-1.004\alpha_{lj}+0.418\alpha_{lj}^{3}-0.169\alpha_{lj}^{5})
\end{equation}

Applying electric analogue, the endothelium permeability $K_{end}$ can
be expressed as a sum of permeabilities of parallel pathways:
\begin{equation}
\label{eq:KK}
K_{end}=K_{lj}+K_{nj}
\end{equation}
The permeability of the leaky junctions $K_{lj}$ is dependent on the
fraction of leaky junctions $\phi$ and may be determined using formula
for the permeability of a slit of width $2 w$ multiplied by area
fraction of a leaky junctions:

\begin{equation}
K_{lj}=\frac{w^{2}}{3}\frac{4w\phi}{R_{cell}}
\end{equation}
WSS does not affect the normal junctions, so $K_{nj}$ is a
constant. Therefore, it can be determined by using known data for
$K_{end}(\phi=5\times10^{-4}) = 3.22 \times 10^{-15} mm^2$\cite{vafai2012}, thus: 

\begin{equation}
K_{nj}=K_{end}(\phi=5\times10^{-4})-K_{lj}(\phi=5\times10^{-4})
\end{equation}

Reflection coefficients for LDL transport are different for normal and
leaky junctions. Normal junctions are impermeable for LDL, so the
reflection coefficient $\sigma_{nj}=1$. Expression for the overall
reflection coefficient of the endothelium can be calculated from
$\sigma$ for each of the individual pathways from heteroporous model
\cite{patlak1971theoretical, rippe1994transport, kellen2003transient}.
Similarly to (\ref{eq:KK}) we have:
\begin{equation}
{K_{end}} \sigma_{end}=K_{lj}\sigma_{lj}+K_{nj}\sigma_{nj}
\end{equation}
thus:
\begin{equation}
\sigma_{end}=1-\frac{(1-\sigma_{lj})K_{lj}}{K_{nj}+K_{lj}},
\end{equation}
where
\begin{equation}
    \sigma_{lj}=1-W=1-(1-\frac{3}{2}\alpha_{lj}^{2}+\frac{1}{2}\alpha_{lj}^{3})(1-\frac{1}{3}\alpha_{lj}^{2})
\end{equation}
and W is overall hindrance factor for convection\cite{silva2009alternative}.

Olgac et al. have applied empirical relations:
(\ref{eq:SI},\ref{eq:MC},\ref{eq:LC},\ref{eq:PHI_WSS}) for
construction of the realistic model comprising the effect of WSS on
LDL transport. We will supplement this approach by taking into the
account the fine structure of the artery wall.

\section{Four-layer model with WSS sensitivity} \label{computer_model}

In this work, we show properties of four-layer arterial wall model,
which includes the impact of WSS on the transport properties given by
the equations (\ref{eq:SI}-\ref{eq:PHI_WSS}).  Our computer simulation
solves equation (\ref{eq:pde2}) in one-dimension, i.e.:
\begin{eqnarray}
   \label{eq:transport1d}
      \frac{\partial c}{\partial t}   & = & -(1-\sigma) u \cdot \frac{\partial c}{\partial x} + D_{eff}\frac{\partial^2 c}{\partial x^2}  - k c
\end{eqnarray} 
with constant filtration velocity calculated from
equation (\ref{eq:filtr}). The equation (\ref{eq:transport1d}) is valid for
coefficients $\sigma,D_{eff}$ independent on space, therefore it is
true in each layer separately. At interfaces between layers the flux
continuity condition must be fulfilled \cite{vafai_yang_2006}:

\begin{eqnarray}
     \left.\left[(1-\sigma)\vec{u}c-D\frac{\partial c}{\partial x}\right]\right|_{+}=\left.\left[(1-\sigma)\vec{u}c-D\frac{\partial c}{\partial x}\right]\right|_{-}
     \label{eq:continuity}
\end{eqnarray}
where $+$ and $-$  particles flux at the the left and right side
of the boundary, respectively.

We solve the equation (\ref{eq:transport1d}) with imposed conditions
(\ref{eq:continuity}) using the finite difference method.  Matrix of
the linear system, generated by the discretization procedure is
sparse, band and symmetrical. We use package {\em sparse} from the
Python SciPy library for the efficient handling and solving our
equations. The details of the implementation as well as the full
source code can be found on git repository \cite{gitLDL}.
 
Results in this work, are based on two sets of transport properties in
the artery: Ai and Vafai (2006) \cite{ai_vafai_2006} presented in the
Table \ref{tab:Ai} (4LA) and Chang and Vafai (2012) \cite{vafai2012}
presented in the Table \ref{tab:Chang} (4LC). 
\begin{table}[t]
	\centering
	\begin{tabular}{|p{3.5 cm}|c|c|c|c|}
	\hline
	\textbf{Names} & Endothelium & Intima & IEL & Media 	\\
	\hline
	\textbf{Thickness }\textit{L, $\mu m$} & $2.0$ & $10.0$  & $2.0$  & $200.0$ \\
	\hline
	\textbf{Diffusivity} \newline \textit{D, $mm ^{2} /s $} & $D(\mathrm{WSS})$& $5.4\times 10^{-6}$ & $3.18 \times 10^{-9}$ & $ 5 \times 10^{-8}$ \\
	\hline
	\textbf{Reflection \newline coefficient} \textit{$\sigma_{1}$} & $\sigma_{1}(\mathrm{WSS})$& $0.8272$  & $0.9827$  & $0.8836$ \\
	\hline
	\textbf{Reaction rate \newline coefficient}  \textit{k, $1/s$} & $0$& $0$  & $0$  & $3.197 \times 10^{-4}$ \\
	\hline
	\textbf{Permeability}  \newline \textit{K, $mm^{2}$} &$K(\mathrm{WSS})$& $2.0\times 10^{-10}  $ &$4.392\times 10^{-13}$&$ 2\times 10^{-12}$  \\
	\hline
	\textbf{Dynamic viscosity}  \textit{$\mu$, $g/(mm\cdot s)$} &$ 0.72\times 10^{-3}  $& $ 0.72\times 10^{-3}  $&$ 0.72\times 10^{-3}  $&$ 0.72\times 10^{-3}  $  \\
	\hline
	\end{tabular}
	\caption{Physiological parameters used in the numerical simulation from Ai and Vafai publication (2006)\cite{ai_vafai_2006} (4LA).  Parameters depending on the WSS are marked by (WSS).}
	\label{tab:Ai}
\end{table}
\begin{table}[t]
	\centering
	\begin{tabular}{|p{3.5 cm}|c|c|c|c|}
	\hline
	\textbf{Names} & Endothelium & Intima & IEL & Media 	\\
	\hline
	\textbf{Thickness }\textit{L, $\mu m$} & $2.0$ & $10.0$  & $2.0$  & $200.0$ \\
	\hline
	\textbf{Diffusivity} \newline \textit{D, $mm ^{2}/s$} & $D(\mathrm{WSS})$& $5.0 \times 10^{-6}$ & $3.18 \times 10^{-9}$ & $ 5 \times 10^{-8}$ \\
	\hline
	\textbf{Reflection \newline coefficient} \textit{$\sigma_{1}$} & $\sigma_{1}(\mathrm{WSS})$& $0.8292$  & $0.8295$  & $0.8660$ \\
	\hline
	\textbf{Reaction rate \newline coefficient}  \textit{k, $1/s$} & $0$& $0$  & $0$  & $1.4 \times 10^{-4}$ \\
	\hline
	\textbf{Permeability}  \newline \textit{K, $mm^{2}$} &$K(\mathrm{WSS})$& $2.2\times 10^{-10}  $ &$3.18\times 10^{-13}$&$ 2\times 10^{-12}$  \\
	\hline
	\textbf{Dynamic viscosity}  \textit{$\mu$, $g/(mm\cdot s)$} &$ 0.72\times 10^{-3}  $& $ 0.72\times 10^{-3}  $&$ 0.72\times 10^{-3}  $&$ 0.72\times 10^{-3}  $  \\
	\hline
	\end{tabular}
	\caption{Physiological parameters used in the numerical simulation from Chung and Vafai (2012) publication  \cite{vafai2012} (4LC). Parameters depending on the WSS are marked by (WSS).}
	\label{tab:Chang}
\end{table}

Additionally, we will also present two-layer model equivalent to that
from Olgac et al. \cite{olgac2008computational}. In that case the
parameters are summarized in Table \ref{tab:2L}.
\begin{table}[t]
\centering
\begin{tabular}{|p{3.5 cm}|c|c|}
\hline
\textbf{Names} & Endothelium & Wall 	\\
\hline
\textbf{Thickness }\textit{L, $\mu m$} & $2.0$ & $338.0$  \\
\hline
\textbf{Diffusivity} \newline \textit{D, $mm^{2}/s$} & $D(\mathrm{WSS})$& $8.0\times 10^{-7}$ \\
\hline
\textbf{Reflection \newline coefficient} \textit{$\sigma_{1}$} & $\sigma_{1}(\mathrm{WSS})$& $0.8514$  \\
\hline
\textbf{Reaction rate \newline coefficient}  \textit{k, $1/s$} & $0$& $3.0 \times 10^{-4}$ \\
\hline
\textbf{Permeability}  \newline \textit{K, $mm^{2}$} &$K(\mathrm{WSS})$& $1.2\times 10^{-12}$\\
\hline
\textbf{Dynamic viscosity}  \textit{$\mu$, $g/(mm\cdot s)$} &$ 0.72\times 10^{-3}  $& $ 0.72\times 10^{-3}  $\\
\hline
\end{tabular}
\caption{Physiological parameters used in the numerical simulation for  two-layer model from Olgac et al. (2006) publication \cite{olgac2008computational}.  Parameters depending on the WSS are marked by (WSS).}
\label{tab:2L}
\end{table}

\section{Results} \label{results}

We solve numerically the equation (\ref{eq:transport1d}) with
Dirichlet boundary conditions: $c(0)=1.0$ and $c(214)=0$. Note that
for clarity we use $\mu m$ as $x$ scale in the numerical algorithm.
The value of the left boundary mimics the concentration of the LDL in
the lumen, therefore due to linearity of the transport equation we
might treat values of $c$ inside the arterial wall as relative
concentrations to the lumen LDL level.  The right boundary is
absorbing, however replacing it with reflecting boundary does not
visibly affect values of $c$ in the $x\in(0,50)$.

The filtration velocity is obtained from (\ref{eq:filtr}) with
effective permeability of the system calculated from the circuit
analogue (see source code \cite{gitLDL}). The transmural pressure
driving the flow in this paper is always taken to be $70\mathrm{mmHg}$.

\subsection{Four-layer model for high WSS} \label{results_4l}

\begin{figure} 
\centering
\includegraphics[width=1.0\textwidth]{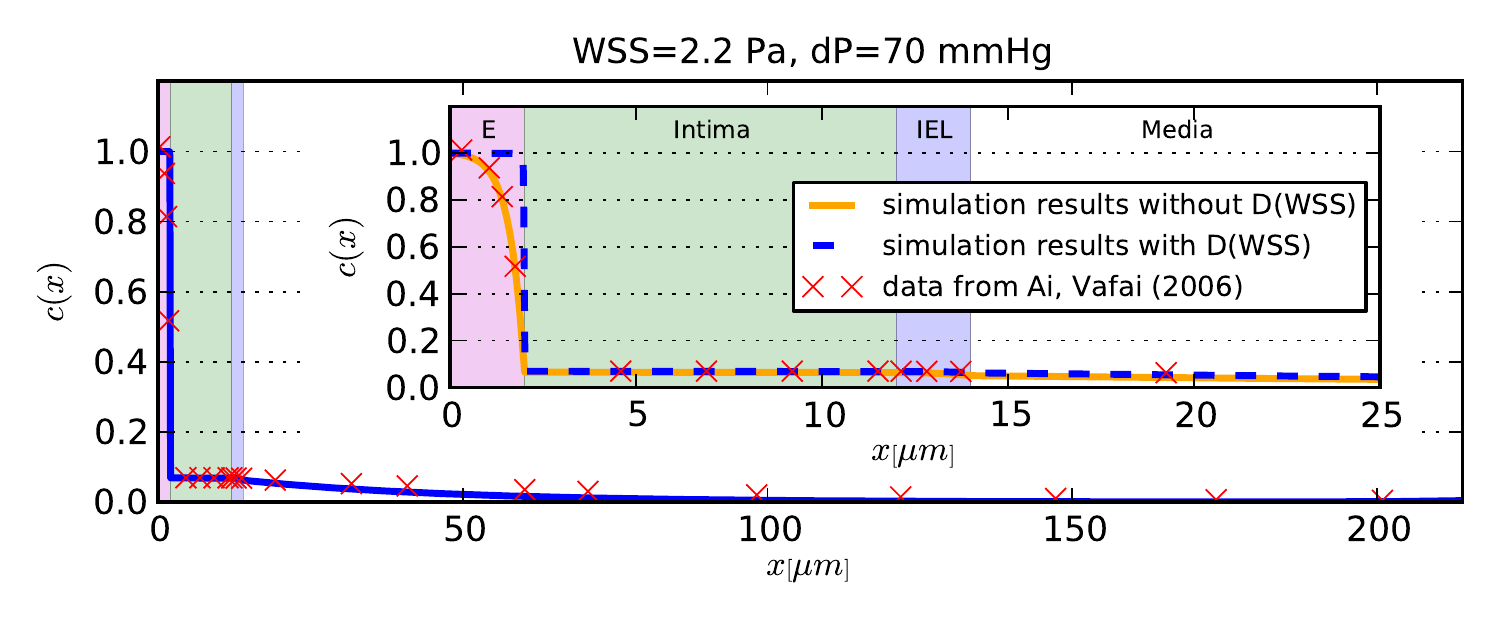}
\caption{ LDL concentration profile in the arterial wall normalized to
  the input concentration $C_0$. $x$ is the distance from the
  lumen. Comparison with the results from the work of Ai, Vafai
  2006. The blue dashed line shows the simulation results for the
  parameters (4LA) from the Table \ref{tab:Ai} and the orange
  continuous line shows the simulation results for the parameters
  (4LA) with the endothelium diffusion coefficient independent on WSS,
  but get from \cite{ai_vafai_2006} equal to $ 8.15 \times 10^{-11}
  mm^2/s $. The red crosses show the data from the work
  \cite{ai_vafai_2006}. Simulation parameters: $dP=70 \mathrm{mmHg}$, $\mathrm{WSS}=2.2$
  Pa. The individual layers are marked with colors: magenta is
  endothelium, green is intima, blue is IEL and white is media. Inset
  graph is an enlargement of the area from $0 \mu m$ to $25 \mu m$.}
\label{fig:AiVafai}
\end{figure}

In the first place our simulations are compared with the results
reported in the literature. Such comparison has been made for two
mentioned sets of parameters. In both cases, the simulations were
performed for WSS equal $2.2 Pa$. The WSS around $2 Pa$, is a typical
physiological value in arteries in free segments without any
obstructions or bifurcations\footnote{We will refer to such value as
  `high`, in contrast to ``low'' value which are present e.g. around a
  stagnation point behind an obstruction.}.  In the Figure
\ref{fig:AiVafai} comparison of the results for the parameters (4LA)
with the data from \cite{ai_vafai_2006} is shown.

An analysis of the graph shows the difference between the obtained
results marked by blue dashed line, and the results from the
publication marked by red crosses. These results slightly differ in
the endothelium concentration profile. To check the reason of this
difference, the parameters, which in our case depend on WSS were
compared with the one used in the \cite{ai_vafai_2006}. That
comparison shows clearly that this difference is caused by the
difference between the diffusion coefficient obtained from the
relation $D(\mathrm{WSS})$ in Equation \ref{eq:D_WSS}, and found in the
publication \cite{ai_vafai_2006}. This is confirmed by the orange
continuous line, which shows results of our simulation with constant
$D(\mathrm{WSS})=8.15 \times 10^{-11} mm^2/s $ coinciding with the
work of Ai and Vafai. An important observation is also that the choice
of the diffusion coefficient does not influence the shape of the
profile in intima, IEL and media layers.

\begin{figure} 
\centering

\includegraphics [width=1.0\textwidth]{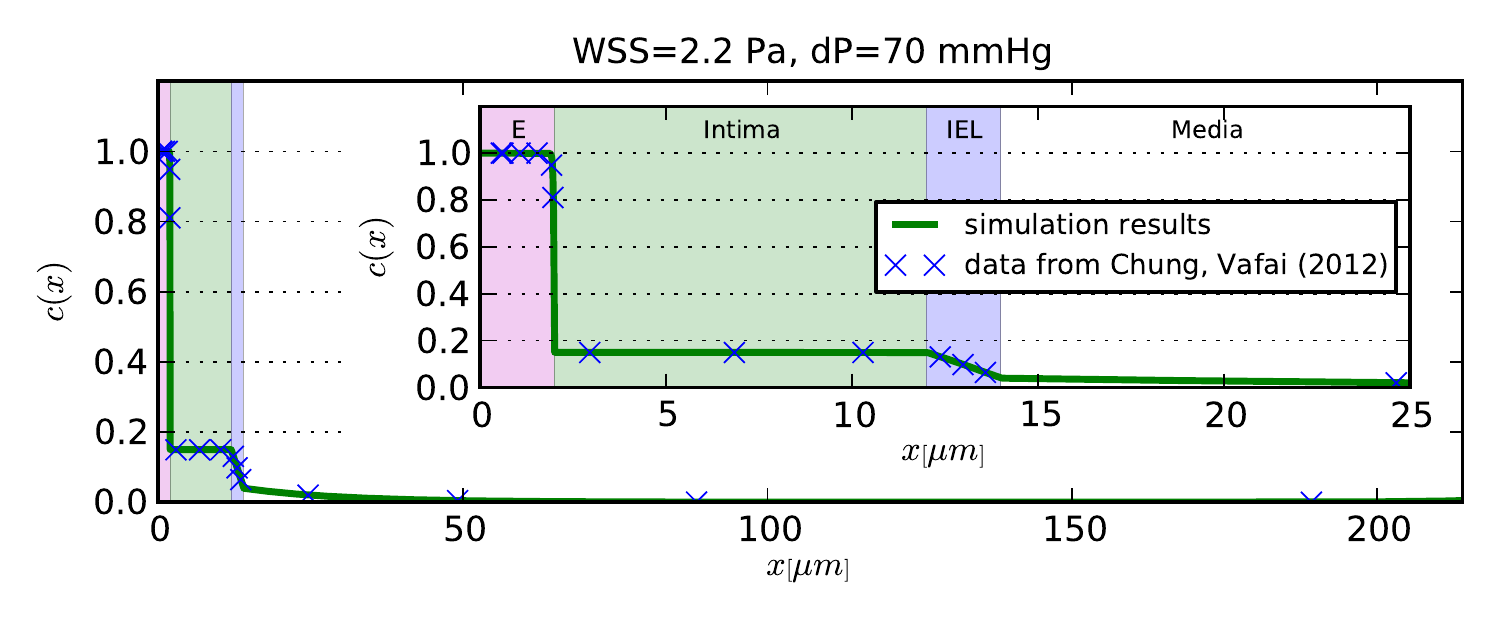}
\caption{LDL concentration profile in the arterial wall normalized to
  the input concentration $C_0$. $x$ is the distance from the lumen.
  Comparison with the results from the work of Chung, Vafai 2012
  \cite{vafai2012}. The green line shows the simulation results for
  the parameters (4LC) from the Table \ref{tab:Chang}. Cyan crosses
  shows data from \cite{vafai2012}. Simulation parameters: $dP=70
  mmHg$, $\mathrm{WSS}=2.2 Pa$. The individual layers are marked with colors:
  magenta is endothelium, green is intima, blue is IEL and white is
  media. Inset graph is an enlargement of the area from $0 \mu m$ to
  $25 \mu m$.}
\label{fig:ChungVafai}
\end{figure}

In the Figure \ref{fig:ChungVafai} a comparison of the results for (4LC)
parameters listed in the Table \ref{tab:Chang} with results from
\cite{vafai2012} is shown. In this case there is no difference in the
endothelium LDL profile. Based on it, it can be assumed that the
dependence $D(\mathrm{WSS})$ from Equation \ref{eq:D_WSS} correctly reproduces the endothelium
diffusion coefficient in function of the WSS.

To sum up, in the case of high WSS that two sets of parameters give
similar LDL concentration profiles.  In this case relationship, known
from the literature, for example from the single-layer model shown in
\cite{olgac2008computational} is visible. The concentration of LDL in
the intima is at the level of one order of magnitude lower then input
concentration (the value depends on the used set of parameters).

\subsection{Giant LDL accumulation effect for low WSS}\label{results_giant}

The most interesting regime is at low WSS values. The low wall shear
stress is present near stagnation points. It can lead to enhanced
penetration of LDL which further is a cause for plague formation in
the arterial wall.

The main barrier in LDL transport at normal physiological conditions
(high WSS) is the endothelium. On the other hand IEL layer has
filtration reflectivity significantly larger than the coefficients of
adjacent layers: the intima and media. If the low WSS influence would
significantly increase transport of LDL through the endothelium, then
it could lead to the accumulation of LDL by the IEL layer. 

\begin{figure} 
\centering

\includegraphics [width=1.0\textwidth]{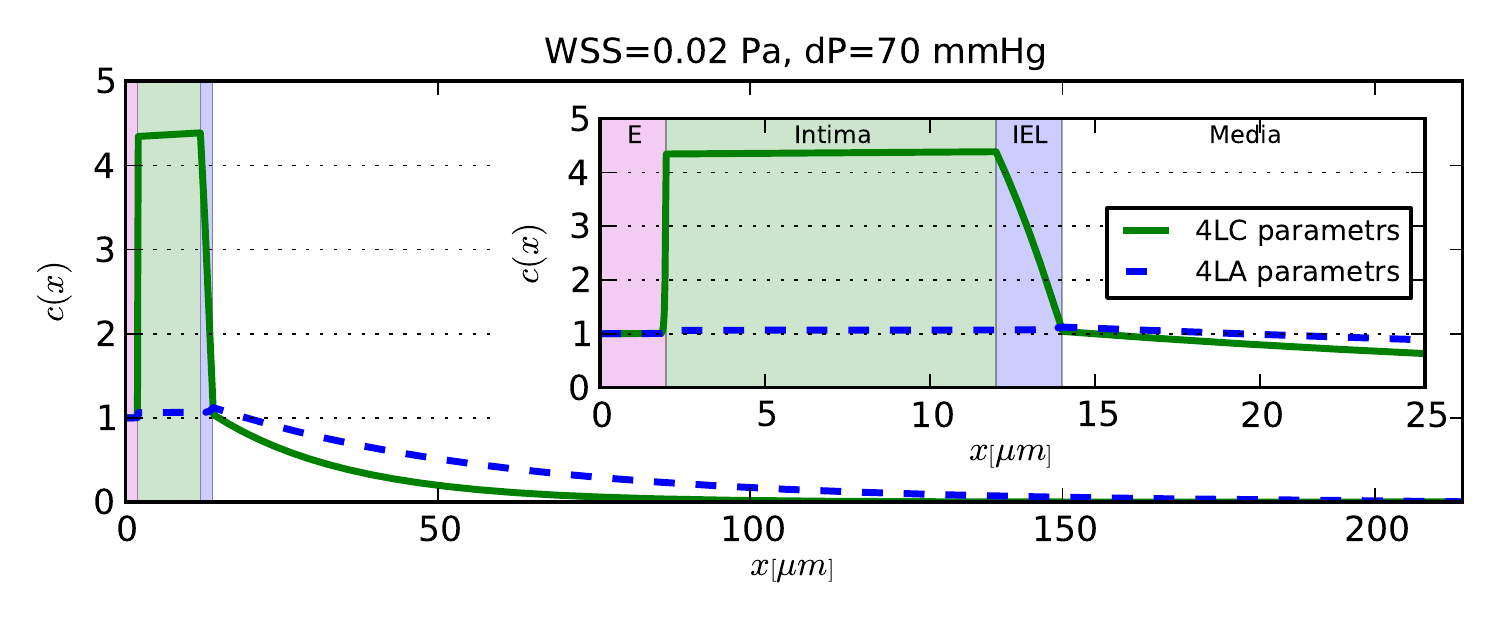}
\caption{LDL concentration profile in the arterial wall normalized to
  the input concentration $C_0$. $x$ is the distance from the
  lumen. Green continuous line shows the simulation results for the
  (4LC) parameters from the Table \ref{tab:Chang}, and the dashed blue
  line shows the simulation results for the (4LA) parameters from the
  Table \ref{tab:Ai}. Simulation parameters: $dP=70 mmHg$, $WSS=0.02
  Pa$. The individual layers are marked with colors: magenta is
  endothelium, green is intima, blue is IEL and white is media. Inset
  graph is an enlargement of the area from $0 \mu m$ to $25 \mu m$.}
\label{fig:WSS002}
\end{figure}

Indeed, this effect can be observed. In Figure \ref{fig:WSS002} we
take very small value of $WSS=0.02 Pa$. The dramatic difference to the
high WSS case is very high level of LDL in the intima. Depending on
the choice of parameters it can even few times exceed the value of the
LDL concentration in the lumen. Note, that this effect is not possible
in the single-layer model.

\begin{figure} 
\centering

\includegraphics [width=1.0\textwidth]{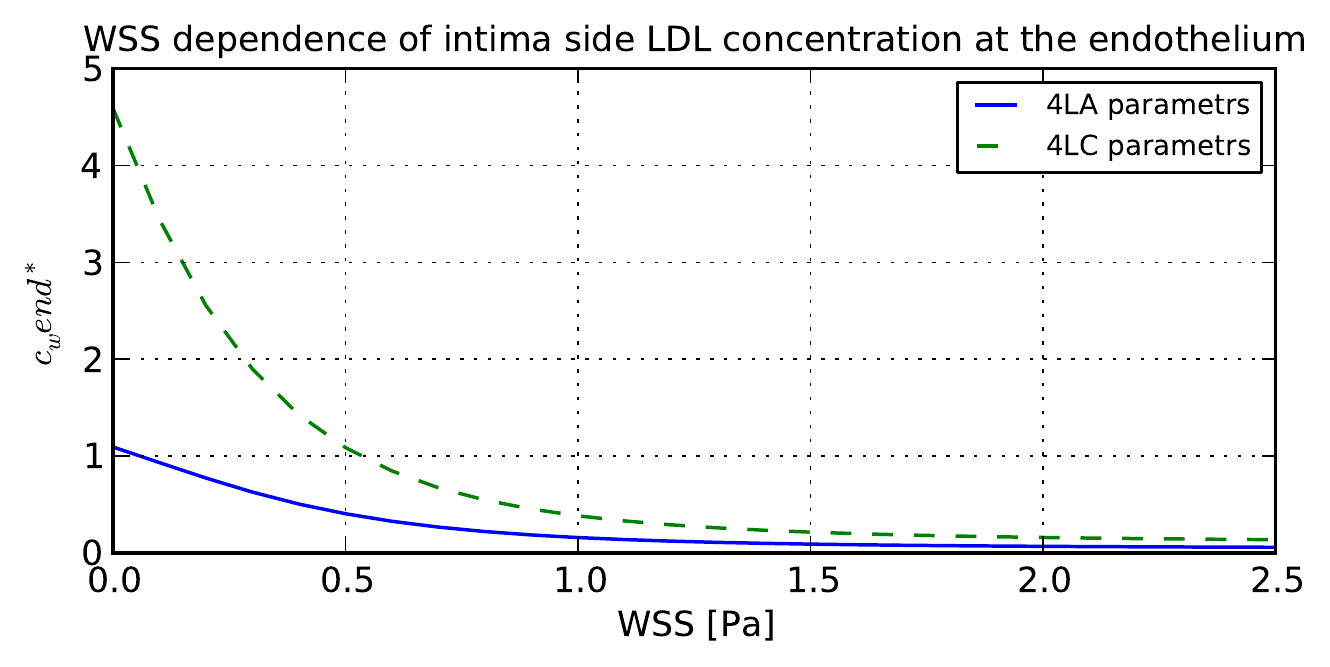}
\caption{WSS dependence of the intima side LDL concentration at the
  endothelium. Green continuous line shows the simulation results for
  the (4LC) parameters from the Table \ref{tab:Chang}, and the blue
  dashed line shows the simulation results for the (4LA) parameters
  from the Table \ref{tab:Ai}.}

\label{fig:4L_WSS}
\end{figure}

The effect of the LDL accumulation in the intima is clearly visible in
the Figure \ref{fig:4L_WSS}, which shows the relationship between WSS
and LDL concentration in the intima on the border of the
endothelium. Despite of the significant difference between the results
for each group of parameters, in both cases the effect of huge LDL
accumulation is visible.

In the single-layer model maximum LDL concentration normalized to
input $C_0$ for the zero WSS is about $0.25$, as shown in the Figure
\ref{fig:2L_Olgac}. In the four-layer model that relative
concentration is $1.0$ or $4.5$, depending on the taken set of parameters,
which is $4 \times$ - $18\times$ greater than in the single-layer
model. It means that the model with internal structure of the arterial
wall predicts the LDL concentrations of the order of magnitude larger
than single-layer one.

\subsection{Properties of two-layer model} \label{results_2l}

We have simplified our LDL transport model to one spatial
dimension. While the arguments of scale separation seem to be valid,
it can be more convincing to present some direct comparisons.

In the work \cite{olgac2008computational} a single-layer model of
blood vessel was used, treating arterial wall as homogeneous layer
with endothelium as a boundary condition. Effectively it means that
the arterial wall consists of the ``zero'' length boundary condition
representing the endothelium and the layer representing the interior
of the wall.  Using our algorithm which is based on transport in
connected regions with piecewiese constant coefficients, we
constructed two-layer model, which effectively corresponds to the
single-layer model used by Olgac et al. In this model, endothelium is
treated as a separate layer (thin, however nonzero length) modeled in
exactly the same way as in the four-layer model with the WSS
dependence. The other layer is considered as a homogeneous porous
medium (arterial wall) with the parameters from the Olgac work. The 
parameters are shown in the Table \ref{tab:2L}.

\begin{figure} 
\centering

\includegraphics [width=1.0\textwidth]{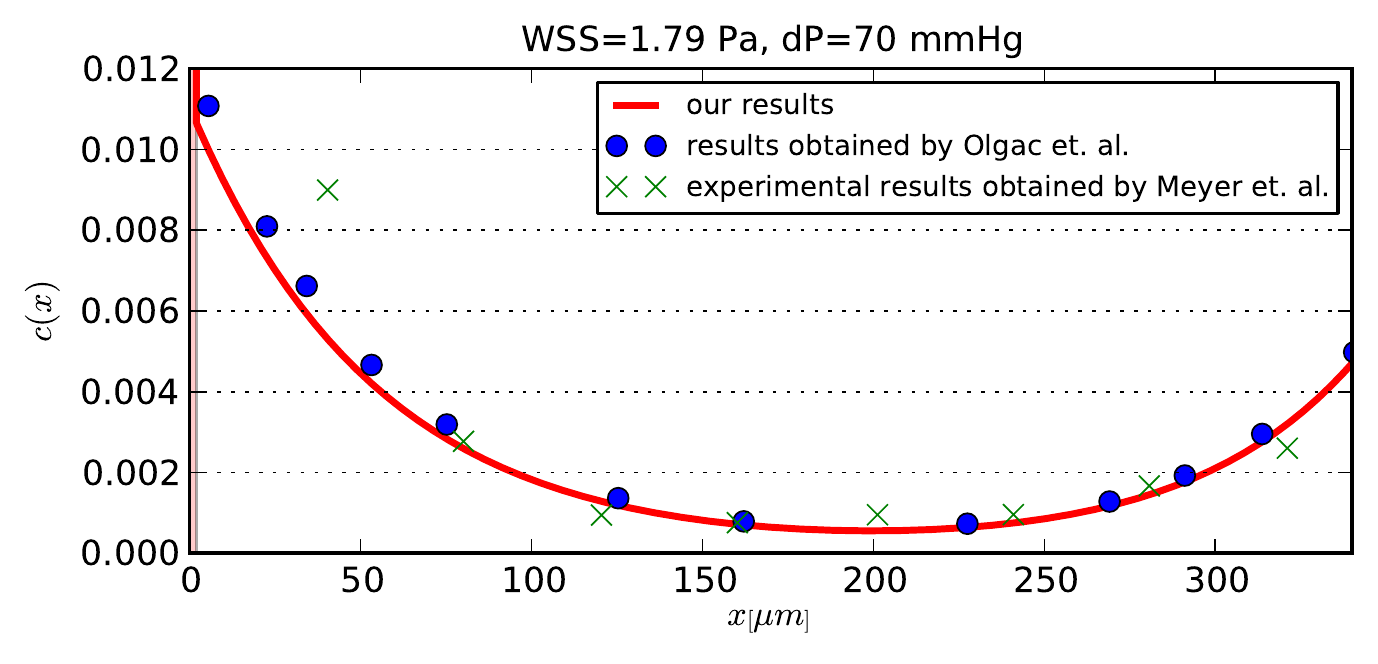}
\caption{Comparison of the computed solute concentration profile
  across the artery wall with the experimental results of Meyer et
  al. \cite{meyer} signed by green crosses and Olgac et al. \cite{olgac2008computational} marked by blue dots at $dP=70 mmHg$ and $WSS=1.79 Pa$. $c$ is the
  arterial wall concentration normalized by the inlet concentration
  $C_0$, and $x$ is the distance from the lumen. The individual layers are marked with colors: red - endothelium and white: rest of arterial wall. In that picture the endothelium layer is almost invisible.}
  
  \label{fig:2L_Olgac}
\end{figure}

In the Figure \ref{fig:2L_Olgac} we show an obtained LDL concentration
profile in the arterial wall together with the computational results
of Olgac et al. from \cite{olgac2008computational} and experimental
results of Meyer et al. from \cite{meyer}. We see that they are
essentially the same. Interestingly, the same concentration profile at
this WSS can be obtained from four-layer model.

\begin{figure} 
\centering

\includegraphics [width=1.0\textwidth]{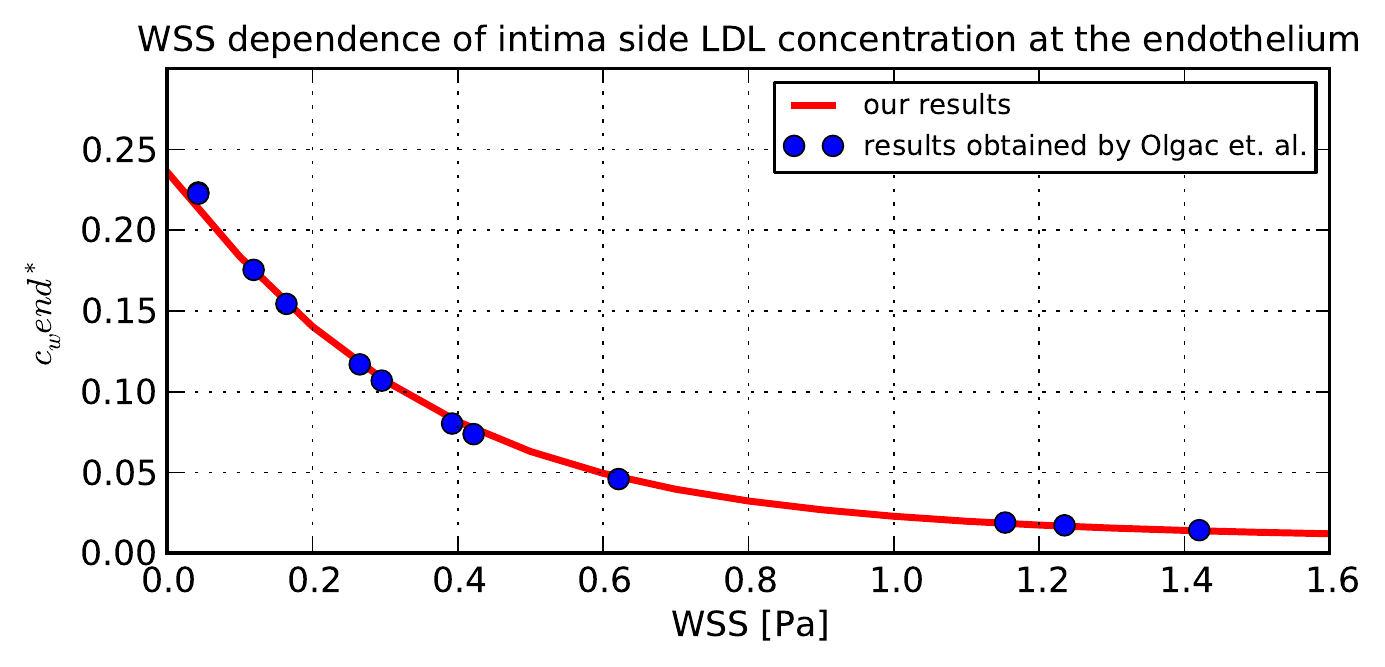}
\caption{WSS Dependence of intima side LDL concentration at the endothelium. Blue line is our simulation results, green crosses are results obtained from paper Olgac et al. \cite{olgac2008computational} }

  \label{fig:2L_WSS}
\end{figure}

The key element of the modeling in this paper is a WSS influence on
the transport. Therefore in Figure \ref{fig:2L_WSS} we compare the
results obtained by Olgac\cite{olgac2008computational} and our
models. We see that results coincide perfectly. This ultimately
provides the evidence that the simplified, one dimensional modeling
can be used in this case. The huge advantage of such simplification is
possibility of exploration of the fine structure of the wall with
relatively small computational effort. Moreover, the single-layer model
cannot predict the effect of giant LDL accumulation.

\section{Conclusions} \label{summary}

The first and main result of this paper is a demonstration of the
giant LDL accumulation in the intima. This effect is related to the
layered structure of the artery. In the four-layer model IEL is a
significant barrier for LDL macromolecules and for this reason they
accumulate in the intima. Our research shows that the LDL
concentration can be up to $18$ times larger than predictions from
single-layer models. We believe that such giant accumulation can play
a significant role in the explanations of the atherosclerosis
development.

Additionally we have shown that the full three dimensional simulation
of the LDL filtration process gives basically the same result as one
dimensional approximation. It can be valuable information for further
construction of advanced models, since it make the three dimensional
solver redundant.

\section{Acknowledgments}
K.J. acknowledges a scholarship from the TWING and FORSZT projects
co-financed by the European Social Fund.

\section{Literature}
\bibliographystyle{elsarticle-num}
\bibliography{paper}
\end{document}